\documentclass[
 ,secnumarabic%
,amssymb,
nobibnotes, aps, prl,
showpacs,showkeys]{revtex4}
\usepackage{DOCS}%
\usepackage{bm}%
\usepackage{graphicx}
\expandafter\ifx\csname package@font\endcsname\relax\else
 \expandafter\expandafter
 \expandafter\usepackage
 \expandafter\expandafter
 \expandafter{\csname package@font\endcsname}%
\fi


\begin{document}

\title{Form Factors in the Quark Resonance Model}%

\author{Krzysztof M. Graczyk}
\email{kgraczyk@ift.uni.wroc.pl}
\author{Jan T. Sobczyk}
\affiliation{ Institute of Theoretical Physics, University of Wroc\l
aw, pl. M. Borna 9, 50-204, Wroc\l aw, Poland}
\date{\today}%

\begin{abstract}
Vector and axial form factors in the quark resonance model are
analyzed with a combination of theoretical and phenomenological
arguments. The new form of form factors is deduced from
$\Delta$(1232) excitation models and available data. The vector
part is shown to agree with the resonant contribution to
electron-proton inclusive $F_2$ data. The axial part is obtained
by finding a simultaneous fit to ANL and BNL $\frac{d \sigma}{d Q^2}$
neutrino scattering data. The best fit corresponds to $C_5^A(0)=0.88$ in the Rarita Schwinger formalism.
\end{abstract}

\maketitle

\section{Introduction}
\label{section1}

New generation of neutrino experiments require better knowledge
about neutrino-nucleon/nucleus cross sections. In the future a lot
of new information will be obtained from the MINER$\nu$A experiment
\cite{MINERVA} but in the meantime one has to rely on the existing
data, theoretical models and information which can be deduced from
electron scattering experiments.

In the 1~GeV neutrino energy region an important contribution to the
total cross section comes from the single pion production (SPP)
channels. The theoretical models which describe SPP reactions are
usually phenomenological in nature and their predictive power is
limited by the precision of SPP neutrino experiments. The standard
description is given in the Rarita Schwinger formalism, with
hadronic current expressed in terms of several form factors
\cite{Adler,Schreiner:1973mj,Fogli:1979cz,Alvarez-Ruso,Lalakulich:2005cs}.
Recently new interesting theoretical approaches were proposed by
Sato and Lee \cite{Sato:2003rq} and Hernandez \textit{et al.}
\cite{Hernandez:2007qq}.

Almost all neutrino interactions Monte Carlo (MC) generators of
events rely on the Rein-Sehgal (RS)
\cite{Ravndal73,Rein:1980wg,Rein:1987cb} model. The RS model is
based on the Feynman-Kislinger-Ravndal (FKR) relativistic quark
model with SU(6) symmetry group \cite{FKR}. It includes
contributions from 18 resonances in the invariant hadronic mass
region $W<2$~GeV. The input to the model consist of: vector and
axial form factors, the value of the Regge slope, masses and widths
of the resonances. The functional forms of vector and axial form
factor were deduced by applying the model to elastic
electron-nucleon and quasi-elastic neutrino-nucleon reactions. The
RS model contains also a prescription how to include a non-resonant
background.

In this paper we propose modifications of the FKR/RS model. They
do not spoil the integrity of the original description and in
particular they leave the same number of free parameters/form
factors. The motivation to our investigation comes from the fact
that in the MiniBooNE and T2K experiments the neutrino beams are
most intensive at the energies $700-800$~MeV. As the consequence
in the inelastic channels the precision of predictions depends
mostly on the quality in which the $\Delta(1232)$ excitation
region is described with higher resonances becoming less
important. This implies that in the RS model the form factors
should be chosen in such a way that $\Delta(1232)$ production is
described as well as possible. In the original FKR/RS model the
form factors are fixed by investigating the elastic and
quasi-elastic reactions. Our choice is to look at $\Delta(1232)$
excitation processes. The advantage of our prescription is that we
obtain form factors which guarantee better description of the
$\Delta(1232)$ excitation region.

In the case of vector form factors we use the recent fits to the
$\Delta$(1232) excitation helicity amplitudes
\cite{Lalakulich:2006sw}. These fits are consistent with the
amplitudes obtained in the MAID model for  electro- and  photo-
production \cite{Drechsel:1998hk}. When applied to the FKR model
some information is lost because in the FKR model the electric
helicity amplitude vanishes. In order to verify our choice we
calculate $F_2$ electron-proton structure function with original and
new vector form factors and conclude that with new form factors the
model is closer to the data. Even better agreement with the data
requires inclusion of background Born terms as it is done in the
MAID model. In our analysis we investigate the resonance form
factors and consequently we focus on the neutrino SPP channel ($\nu
+ p \to \mu^- + p + \pi^+$) in which it is known that the
non-resonance contribution is small. For this reason we find an
agreement with $F_2$ data satisfactory. We did not make a comparison
with electron-neutron data, since they are given in the form of
electron-deuterium data and in  the analysis it is necessary to
eliminate  nuclear effects.

In the case of axial form factor we find simultaneous fits to two
sets (ANL and BNL) of experimental data
\cite{Radecky:1981fn,Kitagaki:1986ct,Kitagaki:1990vs}. We express
our fit for the axial form factor in terms of $C^5_A(Q^2)$ from the
Rarita Schwinger formalism. Then by inverting the reasoning our
results can be interpreted as a fit to $C^5_A(Q^2)$. We consider two
options: with $C^5_A(0)=1.2$ guided by the standard PCAC arguments
or  with $C^5_A(0)$ left as a free parameter. In the second case we
obtain $C^5_A(0)=0.88$ and the agreement with the data is much
better. We notice that recently many authors addressed the problem
of the value of $C^5_A(0)$. In \cite{Hernandez:2007qq} the
introduction of non-resonant background terms in accordance with the
chiral symmetry led authors to the conclusion that the best fit to
both ANL and BNL data is obtained with $C^5_A(0)\approx 0.867$. In
\cite{Alexandrou:2006mc} the lattice QCD results are reported with
$\frac{2C^5_A(0)}{G_A(0)}\approx 1.6$. Computations done in the
chiral constituent quark model reported in
\cite{BarquillaCano:2007yk} give rise to $C^5_A(0) \approx 0.93$.
The main difference between our approach and the one proposed in
\cite{Hernandez:2007qq} is that we do not consider the non-resonant
background. As explained before, we try to avoid the issue of
non-resonant background and we discuss only one SPP channel
$\nu_\mu+p \rightarrow \left(\mu^-+\Delta^{++}\right)\rightarrow
\mu^- +p+\pi^+$ in which it is known that the non-resonant dynamics
is not important. One should remember that above mentioned
evaluations of $C^5_A(0)$ were done under different assumptions
about remaining axial form factors and thus do not necessary mean
the same. For example in \cite{BarquillaCano:2007yk} the authors
obtain $C^3_A(0)\approx 0.035$, $C^4_A(0)\approx -0.25$. The authors
of \cite{Hernandez:2007qq} (as also we do) adopt the Adler model
values: $C^3_A(Q^2)=0$ and $C^4_A(Q^2)=-\frac{C^5_A(Q^2)}{4}$.

The plan of our paper is the following. In Sect.~\ref{section2} we
introduce the basic notation and necessary information about the
objects (helicity amplitudes) calculated in this paper. Sect.~\ref{section4} contains our derivation of
new form factors. The method is based on the analysis of the
existing $\Delta(1232)$ excitation data. Helicity amplitudes are
calculated in two  formalisms which allow to derive
 new RS form factors. Sect.~\ref{section5} contains
comparison of our results with electromagnetic $F_2$ data for
electron-proton scattering and with ANL and BNL neutrino scattering
data. We show  how new form factors modify total cross sections in
charged current (CC) and neutral current (NC) SPP channels.

Throughout this paper we call the discussed model as FKR in the case
of electromagnetic interactions and as RS when it is  applied to
weak interactions.

\begin{figure}[ht]
\centerline{
\includegraphics[width=8cm]{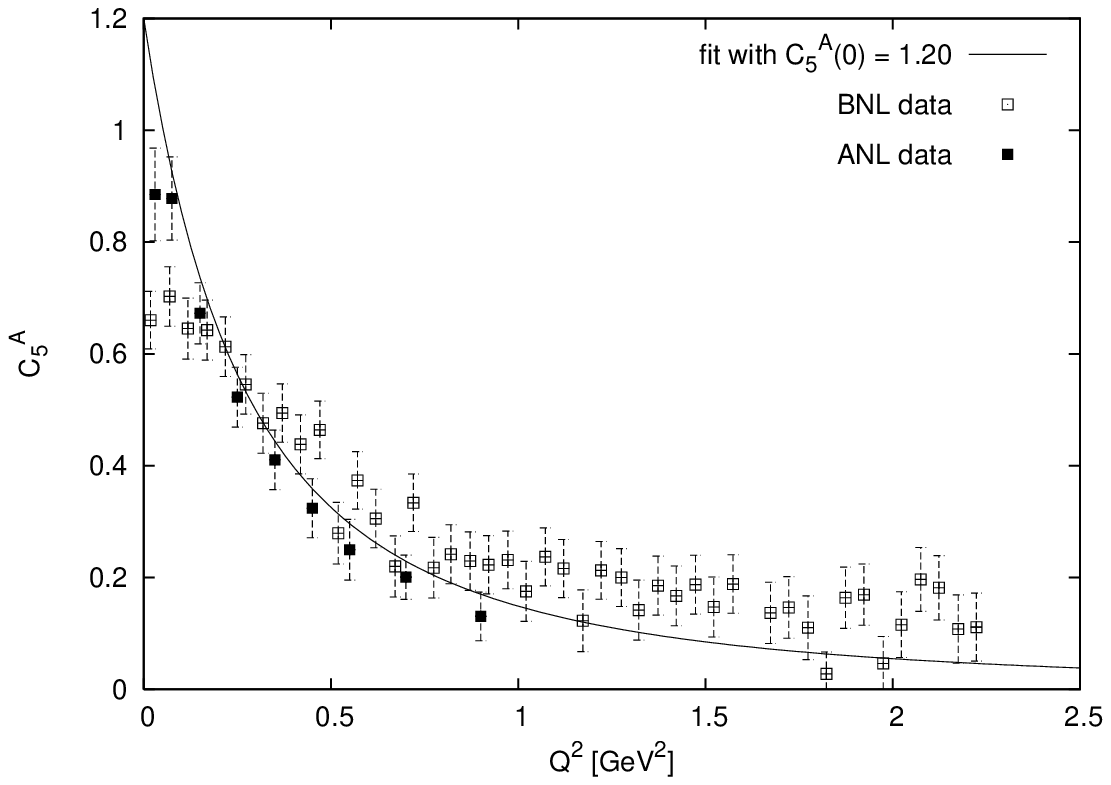}
\includegraphics[width=8cm]{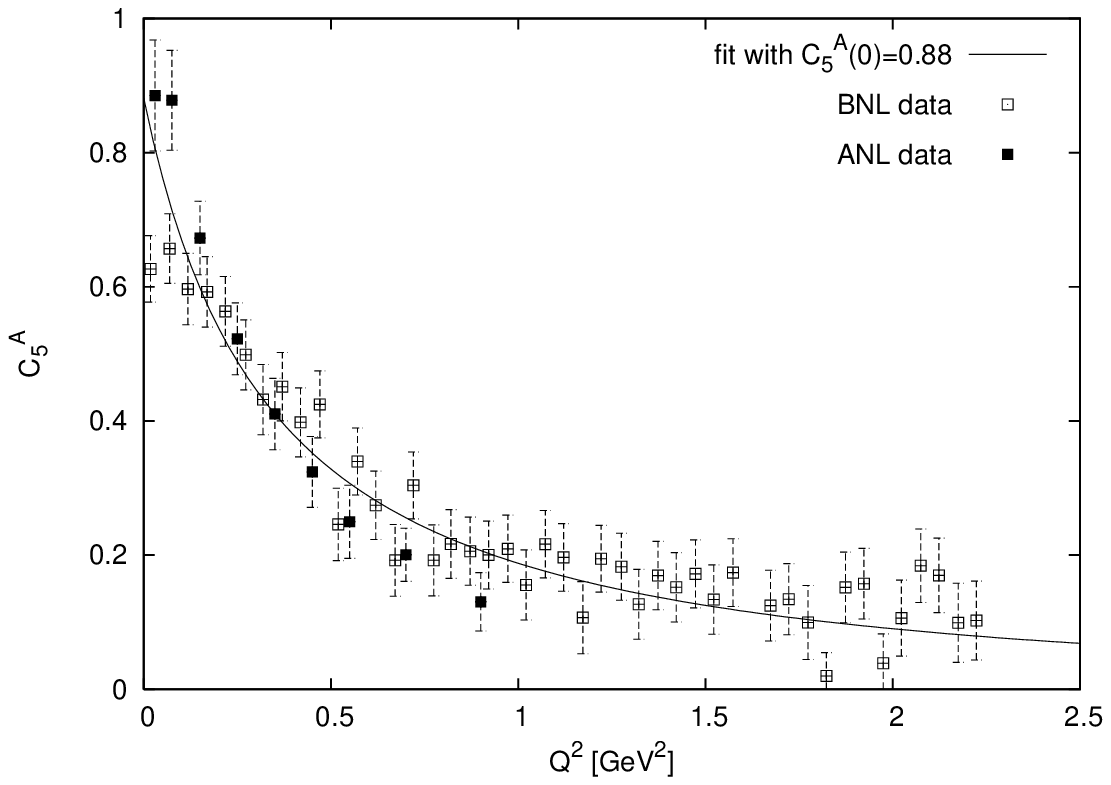}
}\caption{Fits to $C_5^A(Q^2)$ with the ANL (black squares) and the
BNL  (white squares) experimental points. The analytical form of fits are given in Eqs.
(\ref{c5A_fit12})    and  (\ref{c5A_fitbez}). The fitting procedure
is explained in the text.\label{fits_C5A12} }
\end{figure}
\section{Formalism}
\label{section2}

We consider CC neutrino-production of resonances
\begin{equation}
\nu_\mu (k) + N(p)\longrightarrow \mu^- (k') + {\cal N} (p')
\end{equation}
in the framework of the inclusive differential cross section
formalism:
\begin{equation}
{\frac{d^2 \sigma}{d \nu d Q^2}} = \frac{ G^2 \cos^2\theta_C}{ 8
\pi E^2} L_{\mu\nu}W^{\mu\nu},
\end{equation}
where
\begin{equation}
{\rm L}_{\mu\nu} = 2\left({k'}_{\mu}k_\nu + k_{\mu}{k'}_\nu
-g_{\mu\nu}{k'}_\alpha k^\alpha -
i\epsilon_{\mu\nu\alpha\beta}k^\alpha {k'}^\beta  \right)
\end{equation}
and
\begin{equation} W_{\mu\nu} =
(2\pi)^6\sum_{final}\overline{\sum_{s}} \left<final
|\mathcal{J}_\mu^{weak}(0)|p,s\right>
 \left<final|\mathcal{J}_\nu^{weak}(0)|p,s\right>^*\frac{E_p}{M}
\delta^4 \left (   p  + q -  p_{final} \right).
\end{equation}
$k$, $p$, $k'$ and $p'$ denote 4-momenta, the 4-momentum transfer
is: $q^\mu \equiv k^\mu - {k'}^\mu = {p'}^\mu - p^\mu$, $Q^2\equiv
- q_\mu q^\mu$, $k^\mu = (E, \mathbf{k})$, ${k'}^\mu = (E',
\mathbf{k'})$ etc. In the LAB frame the axis orientation is chosen
so that $q^\mu = (\nu, 0, 0, q)$. $M$ denotes the nucleon's and
$M_{R}$'s the resonance mass,
$W$ is the invariant hadronic mass of the final state.

We assume that SPP is mediated by the resonance excitation and we
focus on the computation of independent helicity amplitudes in the
final hadron rest frame:
\begin{eqnarray}
f_{+3} & \equiv & (2\pi )^3 \sqrt{\frac{E_{p,res}}{M}}\left<
{\mathcal N} , p'_{res}, s' =\frac{\displaystyle 3}{\displaystyle
2}\right| \mathcal{J}_+ \left|N,  p_{res}, s=\frac{\displaystyle
1}{\displaystyle
2}\right>,\\
f_{+1}& \equiv & (2\pi )^3 \sqrt{\frac{E_{p,res}}{M}}\left<
{\mathcal N} ,  p'_{res}, s'=\frac{\displaystyle 1}{\displaystyle
2}\right| \mathcal{J}_+ \left|N, p_{res}, s=-\frac{\displaystyle
1}{\displaystyle
2}\right>,\\
 f_{+0}&\equiv & (2\pi )^3 \sqrt{\frac{E_{p,res}}{M}}
 \left< {\mathcal N} ,  p'_{res}, s'=\frac{\displaystyle 1}{\displaystyle 2}\right|
\mathcal{J}_{\underline{0}} \left|N, p_{res}, s=\frac{\displaystyle
1}{\displaystyle 2}\right>.
\end{eqnarray}
$p_{res}$, $s$ and $p'_{res}$, $s'$ denote momenta and spins of
initial ($N$) and final ($\mathcal{N}$) hadrons in the $\mathcal{N}$
rest frame, and $E_{p,res}=\sqrt{M^2+\vec{p}_{res}^2}$.

The definitions of current operators: $\mathcal{J}_+$,
$\mathcal{J}_-$ and $\mathcal{J}_{\underline{0}}$ are \cite{Ravndal73}:
\begin{equation}
\mathcal{J}_\pm =\mp\frac{1}{\sqrt{2}}\left(\mathcal{J}_1 \pm \mathrm{i}\mathcal{J}_2 \right),
\quad \mathcal{J}_{\underline{0}}\equiv \mathcal{J}_0 +
\frac{\nu_{res}}{q_{res}}\mathcal{J}_3.
\end{equation}

Evaluation of the vector part of the current rely on the conserved vector current (CVC)
hypothesis and the comparison with the electromagnetic data is required.
We use the convention in which electromagnetic current is denoted as
$\mathcal{J}_\nu^{em}$ and charged weak current carry no label.
Neutral weak current are discussed only occasionally and then the
label NC is used.

FKR model is a relativistic harmonic oscillator quark model
\cite{FKR}. Resonance wave functions are constructed based on the
SU(6) symmetry \cite{quark_model}. Feynman \textit{et al.}
calculated the hadronic current operators for both electro- and weak
CC neutrino-production of the resonances $\mathcal{J}_\mu^{em}$,
$\mathcal{J}_\mu$. The NC reactions matrix elements are evaluated
with the Standard Model relation \cite{Rein:1980wg}:
\begin{equation}
\mathcal{J}^{NC}_\mu = \mathcal{J}^{CC, I_3}_\mu -  2\sin^2\theta_W
\, \mathcal{J}^{em}_\mu,
\end{equation}
where $\mathcal{J}^{CC, I_3}_\mu$ is a third component of CC
isovector $\mathcal{J}^{CC, I}_\mu$.

In the case of the electro-production the current operators are
multiplied by the an unknown vector form factor $G_V^{RS}$. Similary
the axial part is multiplied by the unknown axial form factor
$G_A^{RS}$:
\begin{equation}
\mathcal{J}_\mu^{em} \to G_V^{RS}(W,Q^2) \mathcal{J}_\mu^{em}, \quad
\mathcal{J}_\mu^{CC}= \mathcal{J}_\mu^{V}- \mathcal{J}_\mu^{A}\to
G_V^{RS}(W,Q^2)\mathcal{J}_\mu^{V} -
G_A^{RS}(W,Q^2)\mathcal{J}_\mu^{A},
\end{equation}

The original way to calculate $G_V^{RS}$ and $G_A^{RS}$ was to
consider elastic electron-nucleon and quasi-elastic neutrino-nucleon
scattering \cite{FKR} (for more details see Appendix A)
\cite{Ravndalthesis,Ravndal72,Ravndal73}. In the vector part the
results are:
\begin{equation}
G_V^{RS}(Q^2) = G_D \left( 1 +
\frac{Q^2}{4M^2}\right)^{
 \frac{1}{2} },\qquad G_D= \left(1 +
\frac{Q^2}{M_V^2}\right)^{-2}.
\end{equation}

The formulas  for the nucleon electric and magnetic form factors
calculated in the FKR/RS model are presented in the Tab.
\ref{tab_1b} Appendix A. We see that they are reproduced in the
approximate way. In the case of proton electric form factor the
difference is  the extra multiplicative factor $(1
-\frac{Q^2}{2M^2})$. In the case of magnetic form factors the proton
and neutron magnetic moments are reproduced with the accuracy of
$\sim 5-7\%$ .

In the FKR/RS model  modifications of $G_V^{RS}$ for
higher level resonances are postulated. New expressions
should lead to the above $G_V^{RS}$ for $N=0$ (level zero in quark
oscillator  model) and $W=M$. The following formula was proposed
in \cite{Ravndal71}:
\begin{equation}
\label{GV_ep} G_V^{RS}(Q^2) = G_D \left( 1 +
\frac{Q^2}{4W^2}\right)^{
 \frac{1-N}{2} }.
\end{equation}
This form factor was used to describe the electro-production data.
To describe the neutrino-production an alternative form was
suggested in \cite{Ravndal73}:
\begin{equation}
\label{GV_nuN} G_V^{RS}(Q^2) = G_D \left( 1 +
\frac{Q^2}{4M^2}\right)^{
 \frac{1}{2}-N}
\end{equation}
which was also adopted in the original RS model. In
\cite{Rein:1987cb} it is explained that the first prescription
(\ref{GV_ep}) is expected to reproduce both resonant and
non-resonant contributions to the inclusive cross section while
the second one (\ref{GV_nuN}) is aimed to describe only the
resonant contribution.

The axial form factor $G_A^{RS}$ is reconstructed from the only one
nonvanishing axial current helicity amplitude (see Tab.
\ref{tab_1c}, Appendix A):
$$
\widetilde G_A^{RS}=ZG_A^{RS}(Q^2)= \frac{3}{5}\left(
1+\frac{Q^2}{4M^2}\right)^{1/2} G_A(Q^2),
$$
where quasi-elastic axial form factor is:
$$
G_{A}(Q^2)= 1.267\left(1 + \frac{Q^2}{M^2_A}\right)^{-2}.
$$

Higher level resonance  modifications are again
postulated \cite{Ravndal72,Ravndal73} and finally:
\begin{equation}
\label{GA_RS}  \widetilde{G}_A^{RS}(Q^2)= 0.76 \left( 1 +
\frac{Q^2}{4M^2}\right)^{ \frac{1}{2}-N } \left(1 +
\frac{Q^2}{M^2_A}\right)^{-2}.
\end{equation}

\section{$\Delta(1232)$ resonance helicity
amplitudes.}

\label{section4}

\subsection{Vector contribution}


The electromagnetic and weak CC excitation of $\Delta(1232)$ can be
modelled using the phenomenological Rarita Schwinger formalism.
\begin{figure}[ht]
\centerline{
\includegraphics{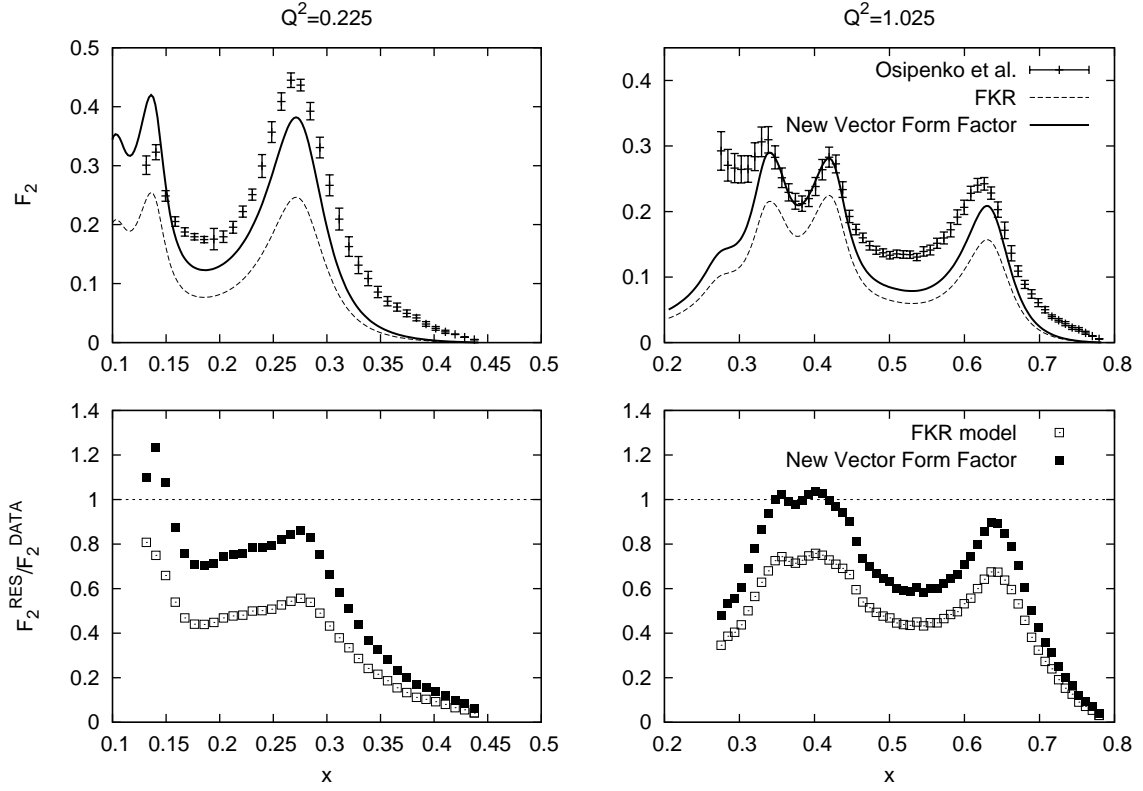}
}\caption{ In the top predictions for $F_2$ for $ep$
scattering in the original FKR model and in the model of this paper
for $Q^2=0.225~\mathrm{GeV}^2$ and $Q^2=1.025~\mathrm{GeV}^2$ are
shown. The data is taken from \protect\cite{Osipenko:2003jb}. In the
bottom  the fractions of the measured strength predicted by both
models are presented. \label{update_F2}}
\end{figure}
The vector part of the charged current (up to normalization it is
also the electromagnetic current) has a general Lorentz covariant
form:
\begin{equation}
\left<\Delta^{++}(p') \right| \mathcal{J}_\mu^{V} \left| N(p)
\right> = \sqrt{3} \bar{\Psi}_\lambda (p') \left[ g^{\lambda}_{\
\mu} T_\nu q^\nu - q^\lambda T_\mu + g^{\lambda}_{\ \mu} C_6^V
\right]\gamma_5 u(p)\frac{1}{(2\pi
)^3}\sqrt{\frac{M_RM}{E_pE_{p'}}},
\end{equation}
where
\begin{equation}
T_\mu =\frac{C_3^V}{M}\gamma_\mu + \frac{C_4^V}{M^2}p'_\mu +
\frac{C_5^V}{M^2}p_\mu,
\end{equation}
$\Psi_\lambda$ is a Rarita-Schwinger field. The conservation
of vector current implies:
\begin{equation}
C_6^V(Q^2) =0.
\end{equation}
We calculate the helicity amplitudes for the $\Delta(1232)$
production in both RS and Rarita Schwinger formalism.

It is enough to consider three independent  amplitudes:
\begin{eqnarray*}
f_{+3}^{\Delta , V}& \equiv & (2\pi )^3
\sqrt{\frac{E_{p,res}}{M}}\left<\Delta , p'_{res}, s'
=\frac{\displaystyle 3}{\displaystyle 2}\right| \mathcal{J}^{V}_+
\left|N, p_{res}, s=\frac{\displaystyle 1}{\displaystyle
2}\right>,\\
f_{+1}^{\Delta , V}& \equiv & (2\pi )^3
\sqrt{\frac{E_{p,res}}{M}}\left<\Delta , p'_{res},
s'=\frac{\displaystyle 1}{\displaystyle 2}\right| \mathcal{J}^{V}_+
\left|N, p_{res}, s=-\frac{\displaystyle 1}{\displaystyle
2}\right>,\\
 f_{+0}^{\Delta , V}&\equiv & (2\pi )^3 \sqrt{\frac{E_{p,res}}{M}}
 \left<\Delta , p'_{res}, s'=\frac{\displaystyle 1}{\displaystyle 2}\right|
\mathcal{J}^{V}_{\underline{0}} \left|N, p_{res},
s=\frac{\displaystyle 1}{\displaystyle 2}\right>.
\end{eqnarray*}

We compare the helicity amplitudes for the vector part of the weak
CC current but the relations we get are the same as in the analysis
of the helicity amplitudes for the electromagnetic current.

In the Rarita Schwinger formalism we obtain:
\begin{eqnarray}
f_{+3}^{\Delta , V} &=&-
N_{q_{res}}\frac{q_{res}}{M+E_{q_{res}}}\left\{\frac{C_4^V}{M^2}
{p'}_\mu q^\mu + \frac{C_5^V}{M^2}p_\mu q^\mu +
\frac{C_3^V}{M}(W+M) \right\}, \\
f_{+1}^{\Delta , V} &=& \sqrt{\frac{1}{3}}
N_{q_{res}}\frac{q_{res}}{M+E_{q_{res}}}
\left\{\frac{C_4^V}{M^2}{p'}_\mu q^\mu + \frac{C_5^V}{M^2}p_\mu
q^\mu + \frac{C_3^V}{M} \left(W+M - 2(M + E_{q_{res}}) \right)
\right\},\\
f_{+0}^{\Delta , V}&=&
-\sqrt{\frac{2}{3}}N_{q_{res}}\frac{q_{res}}{M+E_{q_{res}}}\sqrt{Q^2}
\left\{ \frac{C_4^V}{M^2}W + \frac{C_5^V}{M^2}\frac{M(M+W)}{W}
+\frac{C_3^V}{M}\right\},
\end{eqnarray}
where
\begin{equation}
N_{q_{res}}\equiv \sqrt{\frac{M+E_{q_{res}}}{2M}}, \qquad
E_{q_{res}}=\sqrt{M^2+{q_{res}}^2}.
\end{equation}
The same expressions for the $\Delta(1232)$ helicity amplitudes
were derived before by Lalakulich \textit{et. al}
\cite{Lalakulich:2006sw}.
\\
In the   RS model:
\begin{eqnarray}
f_{+3}^{\Delta , V, RS} &=&-\sqrt{6}\sqrt{\frac{W}{M}}R, \\
f_{+1}^{\Delta , V, RS} &=& -\sqrt{2}\sqrt{\frac{W}{M}}R,\\
f_{+0}^{\Delta , V, RS}&=&0,
\end{eqnarray}
where $$ R\equiv
\sqrt{2}\frac{M}{W}\frac{q(M+W)}{Q^2+(W+M)^2}G_V^{RS}.
$$
The equivalence of both models would mean that:
\begin{eqnarray}
\label{DeltaG3V} G_{V}^{RS}(Q^2,W) & = & \frac{1}{2\sqrt{3}} \left(1
+ \frac{Q^2}{(M+W)^2}\right)^{\frac{1}{2}} \left[C_4^V
\frac{W^2-Q^2-M^2}{2 M^2}+ C_5^V\frac{W^2+Q^2-M^2}{2M^2} +\frac{C_3^V}{M}(W + M) \right],\\
\label{DeltaG1V} G_{V}^{RS}(Q^2,W) & =
&\!\!\!\!-\frac{1}{2\sqrt{3}} \left(1 +
\frac{Q^2}{(M+W)^2}\right)^{\frac{1}{2}}
\left[C_4^V\frac{W^2-Q^2-M^2}{2 M^2} +
C_5^V\frac{W^2+Q^2-M^2}{2M^2} - C_3^V\frac{(M + W)M +Q^2}{MW}
\right],\\
\label{vanishing_V}
0& = &  C_4^V\frac{W}{M^2} +
\frac{C_5^V}{M}\frac{(M+W)}{W} +\frac{C_3^V}{M}.
\end{eqnarray}
In general the above equations with $C_j^V$ provided by experiment
cannot be simultaneously satisfied and solved for $G_{V}^{RS}$. In
particular the quark model predicts that electric contribution  vanishes (Eq. (\ref{vanishing_V})). The well known exception is the
theoretical choice \cite{Schreiner:1973mj}:
\begin{equation}\label{preferred}
C_5^V=0, \quad C_3^V = -\frac{W}{M}C_4^V.
\end{equation}
This preferred from the point of view of the quark model choice is
adopted by many authors. Within this choice there is 1:1
correspondence between $C_4^V$ and $G_V^{RS}$ \cite{Fogli:1979cz}:
$$
C_4^V (Q^2)= -4\sqrt{3}
\left(\frac{M}{M+W}\right)^2  \left(
1+\frac{Q^2}{(M+W)^2}\right)^{-3/2}G_V^{RS}(Q^2).
$$
The problem with the choice (\ref{preferred}) is that it does not
agree well with the  existing electromagnetic data. Our strategy
is to use the fit to the data proposed in
\cite{Lalakulich:2006sw}:
\begin{eqnarray}
C_3^V&=& 2.13 \left( 1 +\frac{Q^2}{4
M_V^2}\right)^{-1}\left(1+\frac{Q^2}{M_V^2} \right)^{-2}, \\
C_4^V&=& -1.51 \left( 1 +\frac{Q^2}{4
M_V^2}\right)^{-1}\left(1+\frac{Q^2}{M_V^2} \right)^{-2}, \\
C_5^V&=&0.48\left( 1 +\frac{Q^2}{4
M_V^2}\right)^{-1}\left(1+\frac{Q^2}{0.776 M_V^2} \right)^{-2}
\end{eqnarray}
with $M_V =0.84$~GeV and translate it into the $G_V^{RS}$.

With such chosen $C_j^V$ we cannot reproduce the quark model
prediction that the electric contribution vanishes and it is clear
that some information has to be lost. In the  Rarita Schwinger
formalism the current is expressed by three functions and in the RS
model by only one. The experimentally measured helicity amplitudes
imply that the significance of  the electric contribution is on the level
of few percent. Since the overall cross section for the pion
production has to be supplemented with a non-resonant contribution
this drawback is not a very serious one.

We notice that the contributions from $f_{+3}$ and $f_{+1}$ enter
the $ep$ cross sections with equal weights. On the other hand, in
the FKR model:
$$
f_{+3}^{\Delta,V, FKR}/f_{+1}^{\Delta,V,FKR} = \sqrt{3}.
$$
Therefore, we propose the following vector form factor:
\begin{equation}
\label{G_V_general_fit} G_V^{RS,new}(W,Q^2) = \frac{1}{2} \sqrt{  3
\left(G^{f_{3}}_{V}(W,Q^2)\right)^2 +
\left(G^{f_1}_{V}(W,Q^2)\right)^2 },
\end{equation}
where
\begin{eqnarray}
\label{G3V} G_{V}^{f_3}(W,Q^2) & \equiv & \frac{1}{2\sqrt{3}}
\left(1 + \frac{Q^2}{(M+W)^2}\right)^{\frac{1}{2}} \left[C_4^V
\frac{W^2-Q^2-M^2}{2 M^2}+ C_5^V\frac{W^2+Q^2-M^2}{2M^2} +\frac{C_3^V}{M}(W + M) \right],\\
\label{G1V} G_{V}^{f_1}(W,Q^2) & \equiv
&\!\!\!-\frac{1}{2\sqrt{3}} \left(1 +
\frac{Q^2}{(M+W)^2}\right)^{\frac{1}{2}}
\left[C_4^V\frac{W^2-Q^2-M^2}{2 M^2} +
C_5^V\frac{W^2+Q^2-M^2}{2M^2} - C_3^V\frac{(M + W)M +Q^2}{M W}
\right].
\end{eqnarray}

We still have to modify  $G_V^{RS,new}$ by a factor describing
modifications of higher resonance excitations. In the expression for
$G_V^{RS}$ (\ref{GV_nuN}) there is the factor $ \left(1
+\frac{Q^2}{4M^2} \right)^{\frac{1}{2}} $ which is obtained from
$$
\lim_{W\to M} \left(1 +\frac{Q^2}{(M+W)^2} \right)^{\frac{1}{2}}.
$$
In the equations (\ref{G3V}) and (\ref{G1V}) there is the same
common factor $ \left(1+\frac{Q^2}{(M+W)^2} \right)^{\frac{1}{2}} $
and it might be natural to keep this term in order to postulate the
higher resonance modification factor. By looking at the duality
properties of the RS model \cite{Graczyk:2005uv} we checked that it
is better to keep this factor the same as in the original FKR/RS
model:
\begin{equation}
\left(1 + \frac{Q^2}{4W^2}\right)^{-\frac{N}{2}}\quad {\rm
or}\qquad \left(1 + \frac{Q^2}{4M^2}\right)^{-N}
\end{equation}
following the arguments presented in the Sect. \ref{section2}.
Therefore we consider two functional forms of the dependence of the
$G_V^{RS,new}$ on the resonance oscillator levels:
\begin{equation}
\label{G_V_general_ep_fit}
 G_V^{RS,new}(W,Q^2) = \frac{1}{2}
 \left(1 + \frac{Q^2}{(M+W)^2}\right)^{\frac{1}{2}}\left(1 + \frac{Q^2}{4W^2}\right)^{-\frac{N}{2}}
 \sqrt{  3 (G_{3}(W,Q^2))^2 +  (G_{1}(W,Q^2))^2 }
\end{equation}
or
\begin{equation}
\label{G_V_general_nu_fit}
 G_V^{RS,new}(W,Q^2) = \frac{1}{2}
 \left(1 + \frac{Q^2}{(M+W)^2}\right)^{\frac{1}{2}}\left(1 + \frac{Q^2}{4M^2}\right)^{-N}
 \sqrt{  3 (G_{3}(W,Q^2))^2 +  (G_{1}(W,Q^2))^2 }
\end{equation}
with
\begin{eqnarray}
G_3(W,Q^2) &=&\frac{1}{2\sqrt{3}} \left[C_4^V \frac{W^2-Q^2-M^2}{2
M^2}+ C_5^V\frac{W^2+Q^2-M^2}{2M^2} +\frac{C_3^V}{M}(W + M)
\right],\\
G_1(W,Q^2) &=&-\frac{1}{2\sqrt{3}} \left[C_4^V\frac{W^2-Q^2-M^2}{2
M^2} + C_5^V\frac{W^2+Q^2-M^2}{2M^2} - C_3^V\frac{(M + W)M +Q^2}{M
W} \right]
\end{eqnarray}
depending on the choice of an ansatz for higher $N$ behavior. We
will use the parametrization (\ref{G_V_general_ep_fit}) for
inclusive $ep$ scattering and (\ref{G_V_general_nu_fit}) for $\nu
N$ scattering in agreement with the logic explained in the Sect.
\ref{section5}.

\subsection{Axial contribution}
\begin{figure}[ht]
\centerline{
\includegraphics[width=19cm]{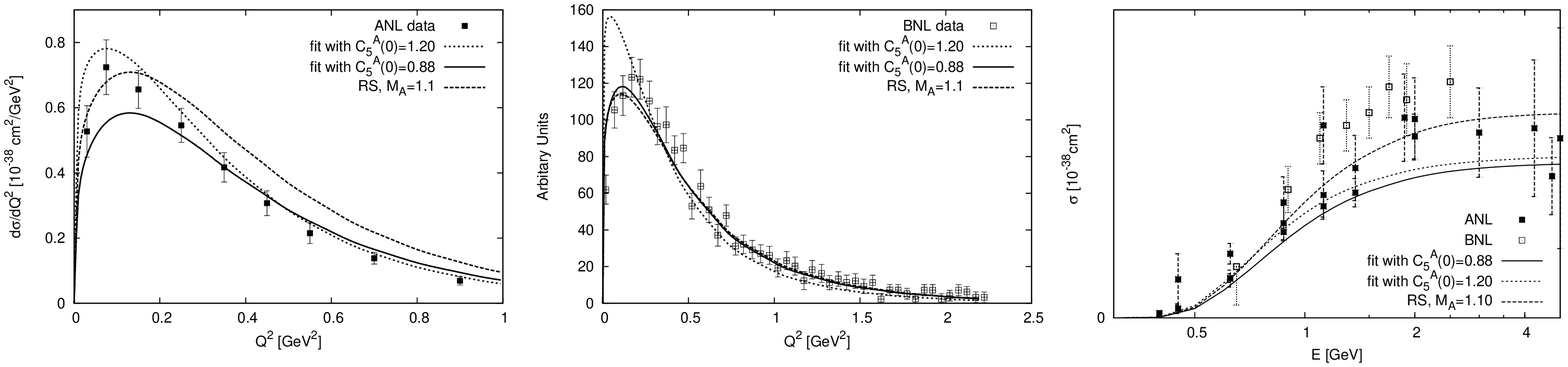}
}\caption{ Differential (left and middle figures)  $d\sigma/dQ^2$
and total cross sections (right figure) for $\nu + p \to \mu^- +p
+\pi^{+}$ scattering. In the first figure the data (black squares)
is from the ANL experiment \protect\cite{Radecky:1981fn}. In the
second figure the data (white squares) is from the BNL experiment
\protect\cite{Kitagaki:1986ct,Kitagaki:1990vs}. Theoretical curves
are obtained with  form factors (\ref{GA_GV_new}) where $C_5^A(Q^2)$
is given by (\ref{c5A_fitbez}) (solid line) and by (\ref{c5A_fit12})
(dotted line). The cross sections  calculated based on original form
factors (\ref{form_factor_set_RS}) with $M_A=1.1$~GeV are denoted by
dashed lines. For  the differential cross sections the cut on the
invariant hadronic mass is imposed $W<1.4$~GeV whereas for the total
cross sections $W<2$~GeV. \label{ANL_Q2_and_BNL_Q2} }
\end{figure}

The axial part of the weak CC current is:
\begin{equation}
\left<\Delta^{++} (p')\right|\mathcal{J}_{\mu}^{A}
\left|N(p)\right> = \sqrt{3}\bar{\Psi}_\lambda (p') \left[
g^{\lambda}_{\ \mu} B_\nu q^\nu - q^\lambda B_\mu + g^{\lambda}_{\
\mu}C_5^A + \frac{q^\lambda q_\mu}{M^2} C_6^A\right]u(p),
\end{equation}
where
\begin{equation} B_\lambda = \frac{C_3^A}{M}\gamma_\lambda +
\frac{C_4^A}{M^2}p'_\lambda.
\end{equation}
The axial contributions to the $\Delta$(1232) helicity amplitudes
are calculated to be:
\begin{eqnarray}
\label{f_A_3} f_{+3}^{\Delta , A} &=& -N_{q_{res}} \left\{
\frac{C_4^A}{M^2}{p'}_\mu q^\mu + C_5^A +
\frac{C_3^A}{M}\left(\nu_{res} + \frac{q_{res}^{2}}{M+E_{q_{res}}}
\right) \right\},
\\
\label{f_A_1} f_{+1}^{\Delta , A} &=& -N_{q_{res}}\sqrt{\frac{1}{3}}
\left\{ \frac{C_4^A}{M^2}{p'}_\mu q^\mu + C_5^A +
\frac{C_3^A}{M}\left(\nu_{res} - \frac{q_{res}^{2}}{M+E_{q_{res}}}
\right) \right\},
\\
\label{f_A_0} f_{+0}^{\Delta , A} &=&
N_{q_{res}}\sqrt{\frac{2}{3}}\left\{-
\frac{\nu_{res}}{\sqrt{Q^2}}C_5^A + \frac{C_3^A}{M}\sqrt{Q^2} +
\frac{C_4^A}{M^2}W \sqrt{Q^2}\right\}.
\end{eqnarray}
In the RS model one obtains:
\begin{eqnarray} \label{f_RS_A_3}
f_{+3}^{\Delta , A, RS} &=&
\sqrt{6}\sqrt{\frac{W}{M}}\frac{\sqrt{2}}{6W}( W+M) \widetilde
G_A^{RS}(W,Q^2),
\\
\label{f_RS_A_1} f_{+1}^{\Delta , A, RS} &=&
\sqrt{2}\sqrt{\frac{W}{M}}\frac{\sqrt{2}}{6W} ( W+M )\widetilde
G_A^{RS}(W,Q^2),
\\
\label{f_RS_A_0} f_{+0}^{\Delta , A, RS} &=& -
2\sqrt{2}\sqrt{\frac{W}{M}}\frac{1}{6Mq} ( W^2 -M^2 )\widetilde
G_A^{RS}(W,Q^2) \frac{q_{res}}{\sqrt{Q^2}}.
\end{eqnarray}
In the comparison we obtain three equations which in general cannot
be simultaneously satisfied.

It is natural to assume that $C_3^A=0$, because then the
relation between $f_{+3}$ and $f_{+1}$ is the same in both
computations:
\begin{equation}
f_{+1}^{\Delta,A} =\sqrt{\frac{1}{3}} f_{+3}^{\Delta,A}.
\end{equation}
In the comprehensive analysis of the $\Delta(1232)$ axial current
\cite{Schreiner:1973mj,Alvarez-Ruso} the following Adler's relation is
assumed:
$$
C_4^A = - \frac{1}{4}C_5^A.
$$
Under this assumption the comparison of axial current helicity
amplitudes leads to equations:
\begin{eqnarray}
\widetilde G_A^{RS,+3, +1}(W,Q^2) &=& \frac{\sqrt{3}}{2}\left(1 +
\frac{Q^2}{(M+W)^2} \right)^{\frac{1}{2}}
\left[1 -\frac{W^2 -Q^2 -M^2}{8M^2} \right] C_5^A(Q^2), \\
\widetilde G_A^{RS,+0}(W,Q^2) &=& \frac{\sqrt{3}}{2}\left(1 +
\frac{Q^2}{(M+W)^2} \right)^{\frac{1}{2}} \left[
\frac{W^2-Q^2-M^2}{2W(W-M)} + \frac{W Q^2}{4M^2(W-M)}
\right]C_5^A(Q^2).
\end{eqnarray}
These are two different expressions  for the $\widetilde
G_A^{RS,new}$. In the cross section the most important region is
that of low $Q^2$ and the difference between them near $Q^2=0$ is
small: $\widetilde G_A^{RS,+3,+1}(W=M_\Delta,Q^2=0)= 0.945$ \newline
and $\widetilde G_A^{RS,+0}(W=M_\Delta,Q^2=0)=0.915$ (we assumed
$C_5^A(0)=1.2$).

We tried to estimate the relative relevance of both amplitudes to
the cross section but it depends on  the neutrino energy and $Q^2$.
We observed also that in the case of $\widetilde G_A^{RS,+0}$ the
increase of the value of $\widetilde G_A^{RS,new}$  with $Q^2$ is
too rapid. In order to be able to get an agreement with both sets of
data we choose:
\begin{equation}
\widetilde G_A^{RS,new}(W,Q^2) = \frac{\sqrt{3}}{2}\left(1 +
\frac{Q^2}{(M+W)^2} \right)^{\frac{1}{2}} \left[1 -\frac{W^2 -Q^2
-M^2}{8M^2} \right] C_5^A(Q^2).
\end{equation}
We see that, under assumptions we have described,  the fit to
$\widetilde G_A^{RS,new}(W,Q^2)$ is equivalent to the fit to
$C_5^A(Q^2)$.

We define an iterative procedure to get $C_5^A(Q^2)$ from the data.
This  procedure takes into consideration  differential cross
sections $\left(\frac{d\sigma}{dQ^2}\right)_{ANL}$ measured in the
ANL experiment \cite{Radecky:1981fn} and the shape of differential
cross section $\left(\frac{d\sigma}{dQ^2}\right)_{BNL}$ measured in
the BNL experiment \cite{Kitagaki:1990vs}. We use also the knowledge
about neutrino fluxes in both experiments.

The fitting procedure consists of several steps:
\begin{enumerate}
\item[(i)] \label{point1} The differential cross section points for
$\left(\frac{d\sigma}{dQ^2}\right)_{ANL}$ are translated into
experimental points for $C^5_A(Q^2)$.
\item[(ii)] The analytical fit to obtained points is found (in order to compare with other approaches
we restricted our analysis to functional forms of $C_5^A(Q^2)$
considered in \cite{Lalakulich:2005cs}).
\item[(iii)] \label{point3} After the obtained fit is used to calculate the flux-averaged cross section with the BNL beam,
the differential cross section points for
$\left(\frac{d\sigma}{dQ^2}\right)_{BNL}$ are translated into
experimental points for $C^5_A(Q^2)$.
\item[(iv)] \label{point4} The simultaneous fit to $C_5^A(Q^2)$ BNL data from point (iii) and  $C_5^A(Q^2)$ ANL data from point (i) is found.
\item[(v)] \label{point5} Using  the new fit from (iv) the steps (iii) and (iv) are repeated.
\end{enumerate}

We define the iterative procedure. The ANL $C_5^A(Q^2)$ points are
unchanged while each iteration moves BNL points. It was checked that
the iterative procedure  is quickly convergent. We needed  four
iteration steps to obtain $C^5_A(Q^2)$ which was virtually unchanged
under further steps. These are the fits discussed in the
remaining part of our paper. In the step (iii) one could have
also started from arbitrary normalization for the BNL cross section. We
checked that our fitting procedure is convergent in this case as well.

We assumed that the relevance of two data sets  is the same. Since
the BNL data consists of  more experimental points we introduced
$\geq 1$ weights to ANL points according to the number od ANL and
BNL points in a given energy bin. Our final fits together with
experimental points extracted from ANL and BNL experiments are shown
in Fig.~\ref{fits_C5A12}. We notice that error bars for the BNL
points for increasing $Q^2$ are quite large. This is because the
relative significance of axial contribution is decreasing.

As explained in the introduction we obtained two fits. In the first
one (case I) we keep the value $C^5_A(0)=1.2$ in accordance with the
PCAC arguments. In the second fit (case II) we treat $C^5_A(0)$ as a
free parameter.

Our results are:
\begin{itemize}
\item case I:
\begin{equation}
\label{c5A_fit12} C_5^A(Q^2) = \frac{C_5^A(0)}{\left( 1 +
\displaystyle\frac{Q^2}{M_a^2}\right)^2 }, \quad C_5^A(0)=1.2, \quad
M_a^2 \approx 0.54\, \mathrm{GeV}^2.
\end{equation}
\item case II:
\begin{equation}
\label{c5A_fitbez} C_5^A(Q^2) = \displaystyle \frac{C_5^A(0)}{\left(
1 + \displaystyle \frac{Q^2}{M_a^2}\right)^2 \left( 1 +
\displaystyle \frac{Q^2}{M_b^2}\right)},\quad C_5^A(0) \approx 0.88,
\quad  M_a^2 \approx 9.71\, \mathrm{GeV}^2, \quad M_b^2 \approx
0.35\, \mathrm{GeV}^2.
\end{equation}
\end{itemize}


Finally, we define the generalization of $\widetilde G_A^{RS,new}$
for higher $N$ along the lines explained before and we obtain:
\begin{equation}
\label{G_A_general_fit} \widetilde G_A^{RS,new}(W,Q^2) =
\frac{\sqrt{3}}{2}\left(1 + \frac{Q^2}{(M+W)^2}
\right)^{\frac{1}{2}} \left(1 + \frac{Q^2}{4M^2} \right)^{-N}\left[1
-\frac{W^2 -Q^2 -M^2}{8M^2} \right]C_5^A(Q^2) .
\end{equation}
\begin{figure}[ht]
\centerline{
\includegraphics{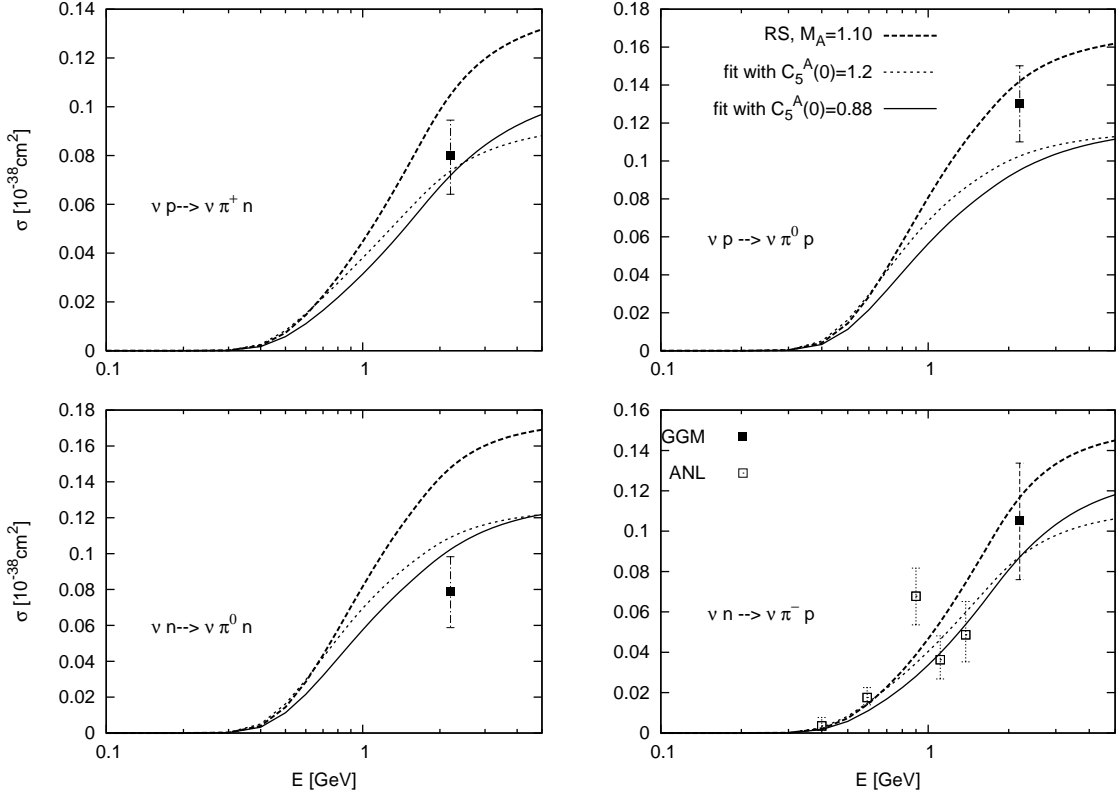}
}\caption{Total cross sections for SPP in NC neutrino-nucleon scattering. The data is from the
experiments: GGM \protect\cite{Krenz:1977sw} (black squares) and ANL
\protect\cite{Derrick:1980nr} (white squares). Theoretical curves are obtained with  form factors (\ref{GA_GV_new}) and
$C_5^A(Q^2)$ given by (\ref{c5A_fitbez})
(solid line) or by (\ref{c5A_fit12}) (dotted line). The cross sections calculated based on
the original RS form factors (\ref{form_factor_set_RS}) with $M_A=1.1$~GeV are denoted by dashed lines. The cut on the invariant hadronic mass $W<2$~GeV is
imposed.\label{total_NC}}
\end{figure}

\begin{table}[ht]
\begin{tabular}{|c|c|c|}
\hline \hline
& & \\
Electromagnetic & & \\ Helicity Amplitudes & Standard Approach & RS model\\
& & \\
\hline
& & \\
$f_{+0}^{em, p}$ & $\displaystyle \left(1+\frac{Q^2}{4M^2}
\right)^{-\frac{1}{2}}G_E^{p}(Q^2)$ & $\displaystyle\left( 1-
\frac{Q^2}{2M^2}\right) \left(1+\frac{Q^2}{4M^2}
\right)^{-1}G_V^{RS}(W,Q^2)$\\
& & \\
& & \\
$f_{+1}^{em, p}$ & $ \displaystyle \frac{q}{\sqrt{2}M}
\left(1+\frac{Q^2}{4M^2}\right)^{-\frac{1}{2}} G_M^{p}(Q^2)$  &
$\displaystyle 3\frac{q}{\sqrt{2}M} \left( 1+\frac{Q^2}{4M^2}\right)^{-1} G_V^{RS}(W,Q^2)$\\
& & \\
& & \\
$f_{+0}^{em, n}$ & $\displaystyle\left(1 +\frac{Q^2}{4M^2}
\right)^{-\frac{1}{2}}G_E^n (Q^2)$ & $0$\\
& & \\
& & \\
$f_{+1}^{em, n}$ & $\displaystyle\frac{q}{\sqrt{2}M}
\left(1+\frac{Q^2}{4M^2}\right)^{-\frac{1}{2}} G_M^{n}(Q^2)$ &
$\displaystyle -2\frac{q}{\sqrt{2}M} \left( 1+
\frac{Q^2}{4M^2}\right)^{-1} G_V^{RS}(W,Q^2)$\\
& & \\
\hline
\end{tabular}
\caption{The elastic electromagnetic helicity amplitudes.
\label{tab_1a}}
\end{table}
\begin{table}
\begin{tabular}{|c|c|c|}
\hline \hline
& & \\
CC Helicity Amplitudes & Standard Approach \cite{llwellyn} & RS model \\
& & \\
\hline
& & \\
$f_{+0}^A$ & $0$  & $0$ \\
& & \\
& & \\
$f_{+1}^A$ & $\displaystyle \sqrt{2}\left(1 +
\frac{Q^2}{4M^2}\right)^{\frac{1}{2}}G_{A}(Q^2)$  & $\displaystyle
\frac{5\sqrt{2}}{3 }ZG_A^{RS}(Q^2)$\\
& & \\
& & \\
$f_{+0}^V$ & $\displaystyle\left(1 +\frac{Q^2}{4M^2}
\right)^{-\frac{1}{2}}\left(G_E^p(Q^2) - G_E^n (Q^2)\right)$  &
$\displaystyle \left( 1- \frac{Q^2}{2M^2}\right)\left( 1+\frac{Q^2}{4M^2}\right)^{-1} G_V^{RS}(Q^2)$ \\
& & \\
& & \\
$f_{+1}^V$ &  $ \displaystyle \frac{q}{\sqrt{2}M}
\left(1+\frac{Q^2}{4M^2}\right)^{-\frac{1}{2}} \left(G_M^p(Q^2)
-G_M^n(Q^2)\right) $  & $\displaystyle
5\frac{q}{\sqrt{2}M}\left(1+\frac{Q^2}{4M^2} \right)^{-{1}}
G_V^{RS}(Q^2)$\\
& & \\
\hline
\end{tabular}
\caption{The quasi-elastic weak CC helicity amplitudes.
\label{tab_1c}}
\end{table}
\begin{center}
\begin{table}[ht]
\begin{tabular}{|c|c|c|}
\hline \hline
 Form Factors & Proton & Neutron \\
  \hline
  &                 &           \\
$G_E(Q^2)$  &    $\displaystyle \left( 1
-\frac{Q^2}{2M^2}\right)\left(1 +
\frac{Q^2}{M_V^2}\right)^{-2}$  &  $0$  \\
  &                 &           \\
 \hline
  &                 &            \\
$G_M(Q^2)$  &    $ \displaystyle 3 \left(1 +
\frac{Q^2}{M_V^2}\right)^{-2}$   & $\displaystyle -2 \left(1 +
\frac{Q^2}{M_V^2}\right)^{-2}
$ \\
  &                 &           \\
  \hline
  \multicolumn{3}{}{}           \\
  \hline\hline
  \multicolumn{3}{|c|}{Axial form factor}             \\
\hline
  &  \multicolumn{2}{|c|}{}           \\
  $G_A(Q^2)$             & \multicolumn{2}{|c|}{$\displaystyle
\frac{5}{3 }Z \left(1 + \frac{Q^2}{M_A^2}\right)^{-2}$}\\
  &  \multicolumn{2}{|c|}{}           \\
\hline \hline

\end{tabular}
\caption{In the top proton and neutron electric and magnetic elastic
form factors obtained within the RS model are shown. In the bottom
the axial nucleon form factor obtained within the RS model is
presented $\left(\frac{5}{3}Z \approx 1.267 \right)$. \label{tab_1b}
}
\end{table}
\end{center}

\section{Discussion}
\label{section5}


In Fig. \ref{update_F2} (top plots) we show predictions of the FKR
model for the electroproduction. In this case the precise data exist
for the inclusive $F_2$ proton structure function
\cite{Osipenko:2003jb}. In the theoretical computation contributions
from 18 resonances (taken form  \cite{Rein:1980wg}) are calculated.
\\
We compare predictions based on the following parameterizations of
$G_V$:
\begin{equation}
G_V^{RS}(W,Q^2) = \left(1 + \frac{Q^2}{4W^2}\right)^{\frac{1-N}{2}}
\left(1 + \frac{Q^2}{M_V^2}\right)^{-2}
\end{equation}
and
\begin{equation}
 G_V^{RS,new}(W,Q^2) = \frac{1}{2}
 \left(1 + \frac{Q^2}{(M+W)^2}\right)^{\frac{1}{2}}
  \left(1 + \frac{Q^2}{4W^2}\right)^{-\frac{N}{2}}
 \left[  3 (G_{3}(W,Q^2))^2 +  (G_{1}(WQ^2))^2 \right]^\frac{1}{2}.
\end{equation}
It is seen that for both values $Q^2=0.225$~GeV$^2$ and
$Q^2=1.025$~GeV$^2$ the results with the new vector form factor are
closer to experimental data. The large difference is seen in
particular in the $\Delta(1232)$ resonance region.

At the $\Delta$(1232) resonance peak some strength is missing, also
with the new form factors, and a non-resonant dynamics is believed
to be responsible for that. In the Fig. \ref{update_F2} (bottom
figures) we show the evaluation of the ratio of the proton $F_2$
calculated within the FKR model (only resonance contribution) and
the experimental data. The computations are done for both form
factors. At the $\Delta(1232)$ resonance peak  with the new vector
form factor the missing strength is $10\div20\%$ depending on the
value of $Q^2$. The similar relative contribution (about 25\%) of
the background dynamics is seen also in plots presented in
\cite{Nakamura:2007pj}.


In CC neutrino-production of resonances vector and axial parts of
the weak current are tested simultaneously. We compare predictions
based on two different sets of form factors. In the first one:
\begin{equation}
\label{form_factor_set_RS}
G_V^{RS} = \left(1 + \frac{Q^2}{4M^2}\right)^{\frac{1}{2}-N} \left(1
+ \frac{Q^2}{M_V^2}\right)^{-2}, \quad \widetilde G_A^{RS}(Q^2) =
0.76 \left( 1 + \frac{Q^2}{4M^2}\right)^{ \frac{1}{2}-N } \left(1 +
\frac{Q^2}{M^2_A}\right)^{-2}
\end{equation}
and in the second one:
\begin{eqnarray}
G_V^{RS,new}(W,Q^2) &=& \frac{1}{2} \left(1 +
\frac{Q^2}{(M+W)^2}\right)^{\frac{1}{2}} \left(1 +
\frac{Q^2}{4M^2}\right)^{-N} \left[  3 (G_{3}(W,Q^2))^2 +
(G_{1}(W,Q^2))^2 \right]^\frac{1}{2}.\nonumber \\
\label{GA_GV_new} \\
\nonumber
\widetilde G_A^{RS,new}(W,Q^2) &=& \frac{\sqrt{3}}{2}\left(1 +
\frac{Q^2}{(M+W)^2} \right)^{\frac{1}{2}} \left(1 + \frac{Q^2}{4M^2}
\right)^{-N}\left[1 -\frac{W^2 -Q^2 -M^2}{8M^2} \right]C_5^A(Q^2)  .
\end{eqnarray}
The first set was used in the original RS paper. As was shown in Sect.
\ref{section4} according to the logic of the RS model $M_A$ should
be the axial mass parameter of the quasi-elastic neutrino
scattering. But usually $M_A$ is considered as a free parameter
fitted with the help of neutrino SPP data. The measurements of $M_A$
give values around 1.00~GeV \cite{Axial_Budd:2004bp}. However,
recent K2K \cite{Axial_mass_Gran:2006jn} and MiniBooNE
investigations \cite{MiniBoone_axial_mass:2007ru} indicate that the
value of $M_A$ can be as big as $1.2$~GeV.  In this paper we show
the predictions of the RS model with the axial mass $M_A=1.1$~GeV
\cite{Furuno:2003ng}.

In the computations with the original RS form factors we take into
account the normalization factors $\mathcal{C}_{\mathcal{N}^*}$
introduced in \cite{Rein:1980wg}  coming from  the Breit-Wigner
amplitudes:
\begin{equation}
\delta(W-M_R) \to  \sqrt{\frac{\Gamma(W)}{2 \pi}}\frac{1}{ W-M_R
+\mathrm{i}\Gamma(W)/2}\cdot
\frac{1}{\sqrt{\mathcal{C}_{\mathcal{N}^*}}},
\end{equation}
where
\begin{equation}
\mathcal{C}_{\mathcal{N}^*} \equiv \int_{W_{thr}}^\infty dW
\frac{\Gamma(W)}{2\pi} \frac{1}{(W-M_R)^2 + (\Gamma(W))^2/4}
\end{equation}
and $W_{thr}=M+m_\pi \approx 1.08~\mathrm{GeV}$ is the threshold for
SPP. For the $\Delta(1232)$ resonance: $\mathcal{C}_{\Delta}\approx
0.87$ and for higher resonances $\mathcal{C}_{\mathcal{N}^*}$ range
from 0.75 to 1.30. In computations with new form factors we do not
include $\mathcal{C}_{\mathcal{N}^*}$ because they are not present
in phenomenological Rarita Schwinger formalism for $\Delta$(1232)
excitation \cite{llwellyn}.

In numerical analysis for neutrino-nucleon interaction we use the RS
approach with lepton mass effects as it is described in
\cite{GS_lepton_mass}.

In Fig. \ref{ANL_Q2_and_BNL_Q2} we compare predictions of RS model
with the experimental results for $\frac{d\sigma}{dQ^2}$ and total
cross section for $\nu + p \to  \mu^- +\Delta^{++}(1232)$. This
reaction is most suitable to discuss because the non-resonant
contribution in the $\Delta$(1232) region is small
\cite{Radecky:1981fn}. We use the data from ANL
\cite{Radecky:1981fn} and BNL \cite{Kitagaki:1986ct,Kitagaki:1990vs}
experiments. The ANL energy beam distribution ranges from $0\div
3$~GeV and has a peak at $E \simeq 0.9$~GeV. The BNL energy beam
distribution ranges from $0\div 6$~GeV and the peak is at $E \simeq
1.2$~GeV. In the case of ANL data the differential cross section is
normalized to the actual cross section and the BNL data are given in
arbitrary units so that only the shape of $\frac{d \sigma}{d Q^2}$
is relevant.

We see that predictions of our model with $C_5^A(0)=0.88$ agree well
with both sets of points. The model with $C_5^A(0)=1.2$ agrees with
ANL data but overestimates  BNL data at low $Q^2$.

We investigated also  the relevance of new form factors for the
prediction of cross sections for NC single pion production (see
Fig.~\ref{total_NC}). In this case only few experimental points
exist. The modification of the form factors  changes the predictions
of the RS model in the  significant way.


\section{Conclusions}

\label{section6}

We proposed new vector and axial form factors which should improve
the performance of the RS model in the $\Delta(1232)$ resonance region. In
the case of axial form factor we consider a simultaneous fit to both
ANL and BNL sets of data without introduction of background terms. Our best fit corresponds to $C_5^A(0)\approx 0.88$. Our results are based on  assumptions specific for the RS model and it
would be interesting to check if the same can be done in the Rarita
Schwinger formalism. Before it was claimed that separate fits must be applied to agree with either ANL or BNL data \cite{Lalakulich:2005cs}.

\section*{Acknowledgements }
\begin{acknowledgements}

The authors were supported by the KBN grant 3735/H03/2006/31. JTS thanks Olga Lalakulich for an
information about the paper \cite{Osipenko:2003jb}.
\end{acknowledgements}


\appendix

\section{}

The quantities to calculate are helicity amplitudes:
\begin{eqnarray*}
f_{+1}^{em,\ N} \equiv  (2\pi )^3 \sqrt{\frac{E_{p,res}}{M}}\left< N
, s'=\frac{\displaystyle 1}{\displaystyle 2}\right|
\mathcal{J}_+^{em} \left|N, s=-\frac{\displaystyle 1}{\displaystyle
2}\right>,\quad
 f_{+0}^{em,\ N} \equiv  (2\pi )^3 \sqrt{\frac{E_{p,res}}{M}}
 \left< N , s'=\frac{\displaystyle 1}{\displaystyle 2}\right|
\mathcal{J}_{\underline{0}}^{em} \left|N, s=\frac{\displaystyle
1}{\displaystyle 2}\right>,
\end{eqnarray*}
where N=$p$ or $n$  denotes nucleon target.

For the CC neutrino-nucleon scattering we need to compute vector and axial
transition matrix elements:
\begin{eqnarray*}
f_{+1}^{V,A} \equiv  (2\pi )^3 \sqrt{\frac{E_{p,res}}{M}}\left< p ,
s'=\frac{\displaystyle 1}{\displaystyle 2}\right|
\mathcal{J}_+^{V,A} \left| n, s=-\frac{\displaystyle
1}{\displaystyle 2}\right>,\quad f_{+0}^{V,A}&\equiv & (2\pi )^3
\sqrt{\frac{E_{p,res}}{M}}
 \left< p , s'=\frac{\displaystyle 1}{\displaystyle 2}\right|
\mathcal{J}_{\underline{0}}^{V,A} \left|n, s=\frac{\displaystyle
1}{\displaystyle 2}\right>.
\end{eqnarray*}

In the case of elastic electron-proton scattering the transition
 matrix elements are:
\begin{eqnarray}
\label{f_mu_em_proton} f^\mu_{em,p}(s', s)
&=& \overline{u}(p',s')\left(F_1^{em,p}(Q^2)\gamma^\mu +
\frac{i\sigma^{\mu\nu}q_\nu}{2 M}F_2^{em,p}(Q^2)  \right) u(p,s).
\end{eqnarray}

$f_\mu$ are computed in the rest frame of the final nucleon. The
Dirac spinors for the incoming and outgoing nucleons are:
\begin{equation}
u(p,s) = \sqrt{\frac{E_{p,res} + M}{2M}}\pmatrix{\displaystyle \chi_s
\cr \displaystyle \frac{\displaystyle -\vec{\sigma}\cdot
\vec{q}_{res}}{\displaystyle E_{p,res} + M} \chi_s},\quad u(p',s') =
\pmatrix{\chi_{s'} \cr 0},
\end{equation}
where $\chi_s$, $\chi_{s'}$  are 2-component spinors.

The relevant combinations of the current (\ref{f_mu_em_proton})
give rise to:
\begin{eqnarray} \label{f_0_em_p} f_{\underline{0}}^{em,p}(s',s)
&=& \chi_{s'}^\dagger G_E^{em,p} \left(1 +\frac{Q^2}{4M^2}
\right)^{-\frac{1}{2}}\chi_s, \\
\label{f_1_em_p} f_\pm^{em,p}(s',s) &=&\mp\frac{1}{\sqrt{2}}(f_1(s',
s) \pm i f_2(s', s))= \chi_{s'}^\dagger\frac{q\sigma_\pm}{\sqrt{2}M}
G_M^{em,p} \left(1+ \frac{Q^2}{4M^2}\right)^{-\frac{1}{2}} \chi_s
\end{eqnarray}
so that
\begin{eqnarray}
f_{+0}^{em,p} &\equiv f_{\underline{0}}(1/2,1/2)  = &G_E^{em,p}
\left(1 +\frac{Q^2}{4M^2} \right)^{-\frac{1}{2}},\\
f_{+1}^{em,p}&\equiv f_{+}(1/2,-1/2)=&\frac{q}{\sqrt{2}M} G_M^{em,p}
\left(1+ \frac{Q^2}{4M^2}\right)^{-\frac{1}{2}}.
\end{eqnarray}

The helicity amplitudes for the electron-neutron scattering are
obtained by substitution  in (\ref{f_0_em_p}-\ref{f_1_em_p})
$G_{E,M}^p \to G_{E,M}^n$.

Similar computations are done for the quasi-elastic CC
neutrino-neutron scattering:
\begin{eqnarray}
f^\mu(s', s) & = & f^\mu_{V}(s',s) - f^\mu_{A}(s',s),
\\
f^\mu_{V}(s',s) &=&\overline{u}(p',s')\left( F_1(Q^2 )\gamma^\mu +
\frac{i\sigma_{\mu\nu}q^\nu}{2 M}F_2(Q^2) \right) u(p,s),\\
f^\mu_{A}(s',s) &=&-\overline{u}(p',s')\left( \gamma^\mu \gamma_5
G_A(Q^2) + q^\mu \gamma_5 F_P(Q^2)\right) u(p,s).
\end{eqnarray}
The vector part of the above current is the same as in the
electromagnetic interactions and to get the matrix elements it is
enough to make a replacement $G_{E,M}^{p} \to G_{E,M}^{p}  -
G_{E,M}^{n} $.

The axial part results are:
\begin{eqnarray}
f_{+0}^{ A}= 0,  \quad f_{+1}^{A}  =  \sqrt{2}\left(1 +
\frac{Q^2}{4M^2}\right)^{\frac{1}{2}}  G_A.
\end{eqnarray}
Analogous calculations, must be done in the Rein-Sehgal model.
Hadronic currents are operators expressed in terms  of spin
$\sigma_a$, isospin $\tau_a$ quark operators and annihilation,
creation ($a$, $a^\dagger$) operators from the 3-dimensional
harmonic oscillator (for detailed explanation see e.g. \cite{FKR}).

The vector components of the hadronic currents read:
\begin{equation}
{\mathcal{J}}_{\underline{0}}^V = 9 \tau_a^+ e^{-\lambda a^{z
\dagger}} S,\quad {\mathcal{J}}_{\pm}^V = 9 \tau_a^+ e^{-\lambda a^{z
\dagger}}\left( T_V a^\dagger_\pm + R_V \sigma^\pm \right),
\end{equation}
where
\begin{equation}
S= \frac{Q^2}{q_{res}^2}\frac{3WM -Q^2 -M^2}{3W}G_V^{RS}, \quad T_V
= \frac{2}{3}\sqrt{\frac{\Omega}{2}} G_V^{RS}, \quad R_V =
\sqrt{2}\frac{2 W q_{res} (M+W)}{(M+W)^2  +Q^2} G_V^{RS}.
\end{equation}
The axial current is expressed as:
\begin{eqnarray}
\mathcal{J}^A_{\underline{0}} = -9 \tau_a^+ e^{-\lambda
a^{z\dagger}}\left(C \sigma^3_a + B
\vec{\sigma}_a\cdot\vec{a}^\dagger \right),\quad \mathcal{J}^A_\pm
&=& \pm 9 \tau_a^\pm e^{-\lambda a^{z\dagger}}\left[R_A \sigma_a^+ +
T_A a_\pm^\dagger \right],
\end{eqnarray}
where
\begin{eqnarray}
B =    G_A^{RS} \frac{2Z}{3} \sqrt{\frac{\Omega}{2}}\left[1 +
\frac{W^2 -Q^2 -M^2}{(W+M)^2 + Q^2} \right], \quad C = G_A^{RS}
\frac{Z W}{3Mq}\left[ W^2 - M^2 + N \Omega \frac{ W^2 -Q^2
-M^2}{(W+M)^2 +Q^2}\right],
\end{eqnarray}
\begin{eqnarray}
R_A = Z G_A^{RS}\frac{\sqrt{2}  }{3} \left(W + M + \frac{2 N \Omega
W}{(W+M)^2 + Q^2}
 \right),
 \quad
T_A =  \frac{4}{3}Z \sqrt{\frac{\Omega}{2}}\frac{q
M}{(W+M)^2+Q^2}G_A^{RS} \sqrt{\frac{\Omega}{2}},
\end{eqnarray}
$\lambda = \sqrt{\frac{2}{\Omega}}q_{res}$, $\Omega =$1.05~GeV$^2$
is determined from the Regge slope of baryon trajectories, $N$ is
the oscillator level of a given resonance.

In the quark model the matrix elements of $\tau_a^+$ and
$\tau_a^+\sigma^\pm_a$ (acting on the first quark) are
\cite{quark_model}:
\begin{eqnarray}
\left< p,\frac{1}{2}\right| \tau_a^+ \left|n,\frac{1}{2}\right> =
\frac{1}{3},\quad \left< p,\frac{1}{2}\right| \tau_a^+
\sigma_+\left|n,-\frac{1}{2}\right> =  \frac{5}{9}.
\end{eqnarray}
Therefore:
\begin{equation}
f_{+0}^{V} = \left(1 + \frac{Q^2}{4M^2}\right)^{-1}\left(1 -
\frac{Q^2}{2M^2}\right)G_V^{RS}, \quad
f_{+1}^{V}=5\frac{q}{\sqrt{2}M}\left(1 + \frac{Q^2}{4M^2}
\right)G_V^{RS} , \quad f_{+0}^{A} = 0 , \quad f_{+1}^{A} =
Z\frac{5\sqrt{2}}{3} G_A^{RS}.
\end{equation}

The outcome of computation is summarized in
Tabs.~\ref{tab_1a}-\ref{tab_1c}, where we collect helicity
amplitudes computed in elastic $eN$ and quasi-elastic $\nu n$
scattering. If we assume that the vector and axial form factors of
the RS model are
\begin{equation}
G_V^{RS}=\left( 1 + \frac{Q^2}{4M^2}\right)^\frac{1}{2}
\left(1+\frac{Q^2}{M^2_V} \right)^{-2}, \quad G_A^{RS}=\left( 1 +
\frac{Q^2}{4M^2}\right)^\frac{1}{2} \left(1+\frac{Q^2}{M^2_A}
\right)^{-2} \end{equation} the electric, magnetic and axial nucleon
form factors take a familiar form shown in Tab. \ref{tab_1b}.

\end{document}